\documentclass[aps, prd, twocolumn, nofootinbib, showpacs]{revtex4}

\usepackage{amsfonts,amsmath,amsthm,amssymb,graphicx,hyperref}
\pdfoutput = 1

\begin{document}

%\preprint{1234}

\title{Attractor Solutions in Tachyacoustic Cosmology}

\author{Dennis Bessada${}^{1,2}$,\footnote{
        {\tt dennis.bessada@unifesp.br},\\
        ${}^\dagger\,{}${\tt whkinney@buffalo.edu}}
        William~H.~Kinney${}^{3,\dagger}$}

     \affiliation{
              ${}^1$Lab. de F\'\i sica Te\'orica e Computa\c c\~ao Cient\'\i fica, Universidade Federal de S\~ao Paulo - UNIFESP, Campus Diadema, Brazil\\
              ${}^2$INPE - Instituto Nacional de Pesquisas Espaciais - Divis\~ao de Astrof\'isica, S\~ao Jos\'e dos Campos, 12227-010 SP, Brazil\\
              ${}^3$Dept. of Physics, University at Buffalo, the  State University of New York, Buffalo, NY 14260-1500, United States}

\begin{abstract}

We study the dynamical stability of ``tachyacoustic'' cosmological models, in which primordial perturbations are generated by a shrinking sound horizon during a period of decelerating expansion. Such models represent a potential alternative to inflationary cosmology, but the phase-space behavior of tachyacoustic solutions has not previously been investigated. We numerically evaluate the dynamics of
two non-canonical Lagrangians, a cuscuton-like Lagrangian and a Dirac-Born-Infeld Lagrangian, which generate a scale-invariant spectrum of perturbations. We show that the power-law background solutions in both cases are dynamical attractors.

\end{abstract}


\maketitle

\section{Introduction}
\label{sec:introduction}

The physics of the very early universe is a rich arena for theory, and there are an abundance
of potential models that solve, at
least partially, the well-known problems of the standard
cosmological paradigm. Simple single-field inflation \cite{inflation} is
surely the most successful model, but alternatives to it have been
proposed: pre-big bang cosmology
\cite{prebigbang}, ekpyrotic and cyclic models \cite{ekpyrotic},
nonsingular quantum cosmological models \cite{quantumcosmology},
non-canonical models \cite{noncanonical}, among others. It was
recently shown that in an expanding cosmology with standard General Relativity,
it is possible to generate a spectrum of super-Hubble curvature
perturbations consistent with data in only three ways \cite{Geshnizjani:2011dk}:
\begin{itemize}
\item{Accelerating expansion ({\it i.e.} inflation),}
\item{A speed of sound faster than the speed of light,}
\item{Super-Planckian energy density.}
\end{itemize}
In this paper we focus on the second of these possibilities, a
superluminal sound speed. In \cite{Bessada:2009ns} we investigated a method of
solving the cosmological horizon problem and seeding
scale-invariant primordial perturbations in a cosmology with {\it
decelerating} expansion and a corresponding {\it growing} comoving
Hubble horizon. If one has a decaying, superluminal sound speed,
curvature perturbations can be generated outside the Hubble
horizon without inflation. We proposed the term {\it
tachyacoustic} for such cosmologies, which are closely related to
varying speed of light theories.

Using a generalization of the inflationary flow formalism
\cite{Kinney:2002qn} introduced by Bean, {\it et al.}
\cite{Bean:2008ga,Kinney:2007ag,Geshnizjani:2011rm} for noncanonical models, we derive two Lagrangians
with solutions which exhibit a shrinking comoving sound horizon
and decelerating expansion. The first is a ``cuscuton'' Lagrangian \cite{Afshordi:2006ad},
which is linear in the scalar field kinetic term, instead of
quadratic as in the case of a canonical Lagrangian. The second is a Dirac-Born-Infeld
(DBI) Lagrangian, of a form similar to that used to implement DBI inflation
in string theory \cite{Silverstein:2003hf,Alishahiha:2004eh}.

The cuscuton Lagrangian is particularly interesting from an observational
perspective. In a recent paper \cite{nongaussian}
it was shown that the cuscuton-like model also leaves a potentially
observable non-Gaussian signature in the CMB anisotropy field,
$f_{\rm NL}\sim \cal{O}$$(1)$, which is slightly different from
the signatures of other superluminal models discussed in the
literature \cite{Magueijo:2010zc,Noller:2011hd}. This difference
is due to the contribution of the term ${\dot{\zeta}^3}$ in the
cubic action \cite{Seery:2005wm,Chen:2006nt} to the full
non-Gaussian amplitude, giving rise to an extra term with a linear
dependence on the flow parameters $\epsilon$ and $s$. This
dependence is not present either in DBI models (since the
coefficient of ${\dot{\zeta}^3}$ in the third-order action is
identically null in this case), or in disformal bimetric model
\cite{Magueijo:2010zc}, since by projecting its scalar-field
action in the Einstein frame one obtains a DBI-like action.
Therefore, the cuscuton tachyacoustic model ``inherits" an extra
contribution from the ${\dot{\zeta}^3}$ term, providing a
different value for $f_{\rm NL}$ compared to the other
superluminal models. Although small, this difference might be
observable in the near future, which is of great importance for
falsifying superluminal noncanonical models.

However, a detailed study of the stability of the dynamical system
for these tachyacoustic models is lacking, and we fill
this gap in the present paper. We investigate the dynamics of both
the cuscuton and DBI Lagrangian, and show that solutions in both cases
are dynamical attractors. This paper is organized as follows. In Section \ref{sec:tachcosm}
we review the basics of tachyacoustic cosmology. In Sec.
\ref{sec:attractors} we discuss the attractor behavior of the
cuscuton-like model (\ref{sec:attcusc}) and the DBI model
(\ref{sec:attdbi}). In Sec. \ref{sec:conc} we summarize the main
results of this paper.

\section{Tachyacoustic Cosmology}
\label{sec:tachcosm}

Tachyacoustic cosmology is a particular solution of a wider class
of k-essence models (see Bean {\it{et al.}} \cite{Bean:2008ga}),
which we briefly review below. Consider a general Lagrangian of
the form ${\cal{L}}={\cal{L}}\left[X,\phi\right]$, where
$2X \equiv g^{\mu\nu}\partial_{\mu}\phi\partial_{\nu}\phi$ is the
canonical kinetic term ($X>0$ according to our choice of the
metric signature). The energy density $\rho$ and pressure $p$ are
given by
\begin{eqnarray}
p &=& {\cal L}\left(X,\phi\right),\label{defP}\\
\rho &=& 2 X {\cal L}_X - {\cal L}\label{defrho},
\end{eqnarray}
The speed of sound is given by
\begin{eqnarray}
\label{defspeedofsound} c_S^{2} &\equiv& \frac{p_X}{\rho_X} =
\left(1 + 2X\frac{{\cal L}_{XX}}{{\cal L}_{X}}\right)^{-1},
\end{eqnarray}
and the corresponding equation of motion for the field $\phi$ is
\begin{equation}
\label{eq:generaleom}
\ddot\phi + 3 H \dot\phi + \frac{{\dot\phi} \partial_0 {\cal L}_X}{{\cal L}_X} - \frac{{\cal L}_\phi}{{\cal L}_X} = 0,
\end{equation}
where the subscript ``${X}$" indicates a derivative with respect
to the kinetic term, and the subscript $\phi$ indicates a derivative with respect to the field. The Hubble parameter $H$ is determined by the Friedmann equation,
\begin{equation}
\label{eqFriedmann} H^2 = \frac{1}{3 M_P^2} \rho = \frac{1}{3
M_P^2} \left(2X {\cal L}_X - {\cal L}\right),
\end{equation}
and the continuity equation is
\begin{equation}
\label{eqcontinuity} \dot\rho = 2 H {\dot H} = -3 H \left(\rho +
p\right) = - 6 H X {\cal L}_X,
\end{equation}
where we use the the reduced Planck mass  $M_P=1/\sqrt{8\pi G}$.
For monotonic field evolution, the field value $\phi$ can be used
as a ``clock'', and all other quantities expressed as functions of
$\phi$, for example $X = X\left(\phi\right)$, ${\cal L} = {\cal
L}\left[X\left(\phi\right),\phi\right]$, and so on. We consider
the homogeneous case, so that $\dot\phi = \sqrt{2 X}$. Next, using
\begin{equation}
\frac{d}{dt} = \dot\phi \frac{d}{d \phi} = \sqrt{2 X}
\frac{d}{d\phi},
\end{equation}
we can re-write the Friedmann and continuity equations as the {\it
Hamilton Jacobi} equations,
\begin{eqnarray}
\label{hamjac1} \dot \phi = \sqrt{2 X} &=& -\frac{2M_P^2}{{\cal
L}_{X}}H'(\phi),\\ \label{hamjac2}
3M_P^2H^2(\phi)&=&\frac{4M_P^4{H'\left(\phi\right)}^2}{{\cal
L}_{X}}-{\cal L}.
\end{eqnarray}
where a prime denotes a derivative with respect to the field
$\phi$. The number of e-folds $N$ is defined as the logarithm of the scale factor,
\begin{equation}
a \propto e^{-N},
\end{equation}
and can be re-written in terms of $d\phi$ by:
\begin{eqnarray}
\label{eq:Nphi} dN = -H dt = - \frac{H}{\sqrt{2 X}} d\phi=
\frac{{\cal L}_X}{2 M_P^2}
\left(\frac{H\left(\phi\right)}{H'\left(\phi\right)}\right) d\phi.
\end{eqnarray}

As in the case of canonical inflation \cite{Kinney:2002qn}, we can
introduce a hierarchy of flow parameters, the first three defined
by
\begin{eqnarray}
\label{defeps} \epsilon\left(\phi\right) &\equiv& \frac{1}{H}\frac{d H}{d N} = \frac{2
M_P^2}{{\cal{L}}_{X}\left(\phi\right)}
\left(\frac{H'\left(\phi\right)}{H\left(\phi\right)}\right)^2,\\
\label{defs} s\left(\phi\right) &\equiv& - \frac{1}{c_S}\frac{d c_S}{d N} = - \frac{2
M_P^2}{{\cal{L}}_{X}\left(\phi\right)}
\frac{H'\left(\phi\right)}{H\left(\phi\right)}
\frac{c_S'\left(\phi\right)}{c_S\left(\phi\right)},\\
\label{defstil} \tilde{s}\left(\phi\right) &\equiv& \frac{1}{{\mathcal L}_X}\frac{d {\mathcal L}_X}{d N} = \frac{2
M_P^2}{{\cal{L}}_{X}\left(\phi\right)}
\frac{H'\left(\phi\right)}{H\left(\phi\right)}\frac{{\cal{L'}}_{X}\left(\phi\right)}{{\cal{L}}_{X}\left(\phi\right)}.
\end{eqnarray}
We construct an exact solution such that the parameters $\epsilon$, $s$, and $\tilde s$ are all identically constant, the background evolution is a power law,
\begin{equation}
a \propto e^{-N} \propto t^{1 / \epsilon},
\end{equation}
and the Hubble parameter and sound speed evolve as
\begin{eqnarray}
\label{plevolution}
&&H \propto e^{\epsilon N}\cr
&&c_S \propto e^{-s N}\cr
&&{\cal L}_{X} \propto e^{{\tilde s} N}.
\end{eqnarray}
In terms of the field $\phi$, this solution to the flow equations corresponds to
\begin{equation}
\label{sqrtlagX}
{\cal{L}}_{X} = \frac{8\epsilon}{{\tilde s}^2}\left(\frac{M_P}{\phi}\right)^2,
\end{equation}
\begin{equation}
\label{hubpar}
H\left(\phi\right)=H_0\left(\frac{\phi}{\phi_0}\right)^{-2\epsilon/\tilde{s}},
\end{equation}
\begin{equation}
\label{gamm}
c_S\left(\phi\right)=\left(\frac{\phi}{\phi_0}\right)^{2s/\tilde{s}},
\end{equation}
where $\phi_0$ is a fiducial field value defined such that $c_S\left(\phi_0\right) = 1$, and the field evolves as
\begin{equation}
\label{efoldsphi}
\frac{\phi}{\phi_0} = e^{-{\tilde s} N / 2}.
\end{equation}
From Eqs. (\ref{defstil},\ref{sqrtlagX}), we see that $\phi_0$ is not an independent constant, but is given by
\begin{equation}
\left(\frac{\phi_0}{M_P}\right)^2 = \frac{8 \epsilon}{{\tilde s}^2}.
\end{equation}
From the Hamilton-Jacobi Equation (\ref{hamjac1}), the field velocity is then
\begin{equation}
\label{generaldotphi}
{\dot\phi} = \frac{{\tilde s} \phi_0 H_0}{2} \left(\frac{\phi}{\phi_0}\right)^{1 - 2 \epsilon / {\tilde s}}.
\end{equation}
We then {\it construct} a Lagrangian for which Eqs. (\ref{sqrtlagX},\ref{hubpar},\ref{gamm}) are an exact solution to the associated equations of motion. The important point is that given a solution to the flow equations, the associated Lagrangian is fully determined, up to the specification of a choice of gauge, {\it i.e.} the relationship between $s$ and $\tilde s$ \cite{Bean:2008ga}. For a particular choice of gauge, we construct the Lagrangian as follows \cite{Bessada:2009ns}: From Eqs. (\ref{sqrtlagX}) and (\ref{gamm}), we see that the speed of sound $c_S$ can be written in terms of ${\cal{L}}_{X}$
\begin{equation}
\label{lagXgamma}
c_S^2 = \left[1 + 2 X \frac{{\cal L}_{XX}}{{\cal L}_X}\right]^{-1} =C^{-1} {\cal{L}}_{X}^{-2 s/\tilde{s}},
\end{equation}
where we have used Eq. (\ref{defspeedofsound}), and defined
\begin{equation}
\label{defK}
C \equiv \left(\frac{\tilde{s}^2\phi_0^2}{8M_P^2\epsilon}\right)^{2 s / \tilde s}.
\end{equation}
The result is a differential equation for the Lagrangian ${\cal{L}}\left(X,\phi\right)$:
\begin{equation}
\label{difflag}
2X{\cal{L}}_{XX}+{\cal{L}}_X-C {\cal{L}}_{X}^{n}=0,
\end{equation}
where we have defined
\begin{equation}
\label{difflagconst}
n\equiv 1 + \frac{2 s}{\tilde{s}}.
\end{equation}
Therefore, by specifying a relationship between the parameters $s$ and $\tilde s$, we can construct a Lagrangian as the solution to the differential
equation (\ref{difflag}). For example, a canonical Lagrangian with speed of sound $c_S = {\rm const.} = 1$ is just the case $s = 0$, so that $n = 1$
and $C = 1$, and Eq. (\ref{difflag}) becomes
\begin{equation}
{\cal{L}}_{XX}=0,
\end{equation}
with general solution
\begin{equation}
{\cal L} = f\left(\phi\right) X - V\left(\phi\right).
\end{equation}
Here $f\left(\phi\right)$ and $V\left(\phi\right)$ are free functions which arise from integration of the second-order equation (\ref{difflag}).
The function $f\left(\phi\right)$ can be eliminated by a field redefinition $d \varphi = \sqrt{f\left(\phi\right)} d \phi$, resulting in a manifestly canonical Lagrangian for $\varphi$, as we would expect from setting $c_S = 1$. We emphasize that Eq. (\ref{difflag}) is constructed using the solution (\ref{gamm}), and is not a general condition on the Lagrangian. That is, Eq. (\ref{difflag}) allows us to construct a Lagrangian which admits solutions of the desired form, but those solutions are not necessarily unique.  A canonical Lagrangian can support inflationary solutions, but not
tachyacoustic solutions, and is therefore not of interest here. However, other choices of $n$ do yield tachyacoustic solutions, and we focus on two
such choices:
\begin{enumerate}
\item{$n=0$: A Cuscuton-like model.}
\item{$n=3$: A DBI model.}
\end{enumerate}
In the next section, we consider the attractor behavior of the solution given by Eqs. (\ref{sqrtlagX},\ref{hubpar},\ref{gamm}) in both the Cuscuton and DBI cases.

\section{Attractor Behavior}
\label{sec:attractors}
\subsection{The Cuscuton Case}
\label{sec:attcusc}

The case where the exponent $n = 0$ in Eq. (\ref{difflag}) corresponds to taking $\tilde{s}=-2s$, and the resulting Lagrangian is \cite{Bessada:2009ns}
\begin{equation}
\label{lagcuscs}
{\cal{L}}\left(X,\phi\right)=2f\left(\phi\right)\sqrt{X} + C X -
V\left(\phi\right),
\end{equation}
where $f\left(\phi\right)$ is an arbitrary function, $V\left(\phi\right)$ is the potential and the constant $C$ in Eq. (\ref{defK}) is given by
\begin{equation}
\label{defC} C=\frac{2M_P^2\epsilon}{s^2\phi_0^2}.
\end{equation}
The power-law solution (\ref{sqrtlagX},\ref{hubpar},\ref{gamm}) corresponds to the choice
\begin{eqnarray}
\label{deff} f\left(\phi\right)&\equiv&
\frac{\sqrt{2} M_P^2 H_0 \epsilon}{s \phi_0} \left(\frac{\phi}{\phi_0}\right)^{\epsilon / s + 1} \left[\left(\frac{\phi_0}{\phi}\right)^2-1\right],
\\
\label{defV}
V\left(\phi\right)&\equiv& 3 M_P^2 H_0^2\left(\frac{\phi}{\phi_0}\right)^{2\epsilon / s} \left[1 -\frac{\epsilon}{3} \left(\frac{\phi}{\phi_0}\right)^2\right],
\end{eqnarray}
The Hubble parameter (\ref{eqFriedmann}) is
\begin{equation}
H = \sqrt{\frac{1}{3 M_P^2} \left[\frac{C}{2} \dot\phi^2 + V\left(\phi\right)\right]},
\end{equation}
and the sound speed (\ref{defspeedofsound}) is
\begin{equation}
c_S = \sqrt{\frac{{\mathcal L}_X}{C}} = \sqrt{1 + \frac{\sqrt{2} f\left(\phi\right)}{C \dot \phi}}.
\end{equation}
The equation of motion for the Lagrangian (\ref{lagcuscs}) is
\begin{equation}
\label{eq:cusceom}
\ddot \phi + 3 H c_S^2 \dot\phi + C^{-1} V'\left(\phi\right) = 0,
\end{equation}
for which the exact solution (\ref{sqrtlagX},\ref{hubpar},\ref{gamm}) is
\begin{eqnarray}
\label{eq:cuscutonsolution}
&&\dot\phi = - s H_0 \phi_0 \left(\frac{\phi}{\phi_0}\right)^{\epsilon / s + 1},\cr
&&H\left(\phi\right) = H_0 \left(\frac{\phi}{\phi_0}\right)^{\epsilon / s},\cr
&&c_S\left(\phi\right)= \sqrt{\frac{{\cal L}_x}{C}}= \left(\frac{\phi_0}{\phi}\right).
\end{eqnarray}
It is straightforward to verify that this solution corresponds to power-law evolution,
\begin{equation}
a \propto t^{1 / \epsilon} \propto e^{-N},
\end{equation}
where the Hubble parameter and speed of sound evolve according to Eq. (\ref{plevolution}).
The resulting model is {\it{tachyacoustic}}, that is, has a
decaying, superluminal speed of sound and shrinking acoustic
horizon in comoving units.

It can be shown \cite{Bessada:2009ns} that such cosmologies produce a power-law spectrum of superhorizon curvature perturbations with spectral index
\begin{equation}
n = 1 - \frac{2 \epsilon + s}{1 - \epsilon - s},
\end{equation}
so that the scale-invariant limit is $s = -2 \epsilon$. A red-tilted scalar spectrum index of
perturbations such as that favored by WMAP, $n = 0.96 \pm 0.026$ \cite{Komatsu:2008hk} can be
achieved if we choose the value of the flow parameter $s$ to lie in the interval $s = [-1.907 \epsilon,-1.980 \epsilon]$. (For the remainder of this paper we will consider the scale invariant limit $s = -2 \epsilon$, which is sufficient for the purpose of demonstrating a dynamical attractor.) In this sense, tachyacoustic cosmology represents an interesting possible alternative to inflation: instead of accelerating expansion, tachyacoustic models produce a spectrum of perturbations consistent with the data via a decreasing, superluminal sound speed in (for example) a matter- or radiation-dominated background. However, in order to represent a viable cosmological solution, the analytic solution (\ref{eq:cuscutonsolution}) must be a stable dynamical attractor, so that many different choices of boundary condition will converge on a single solution at late time.

\begin{figure}
\includegraphics[width=3.0in]{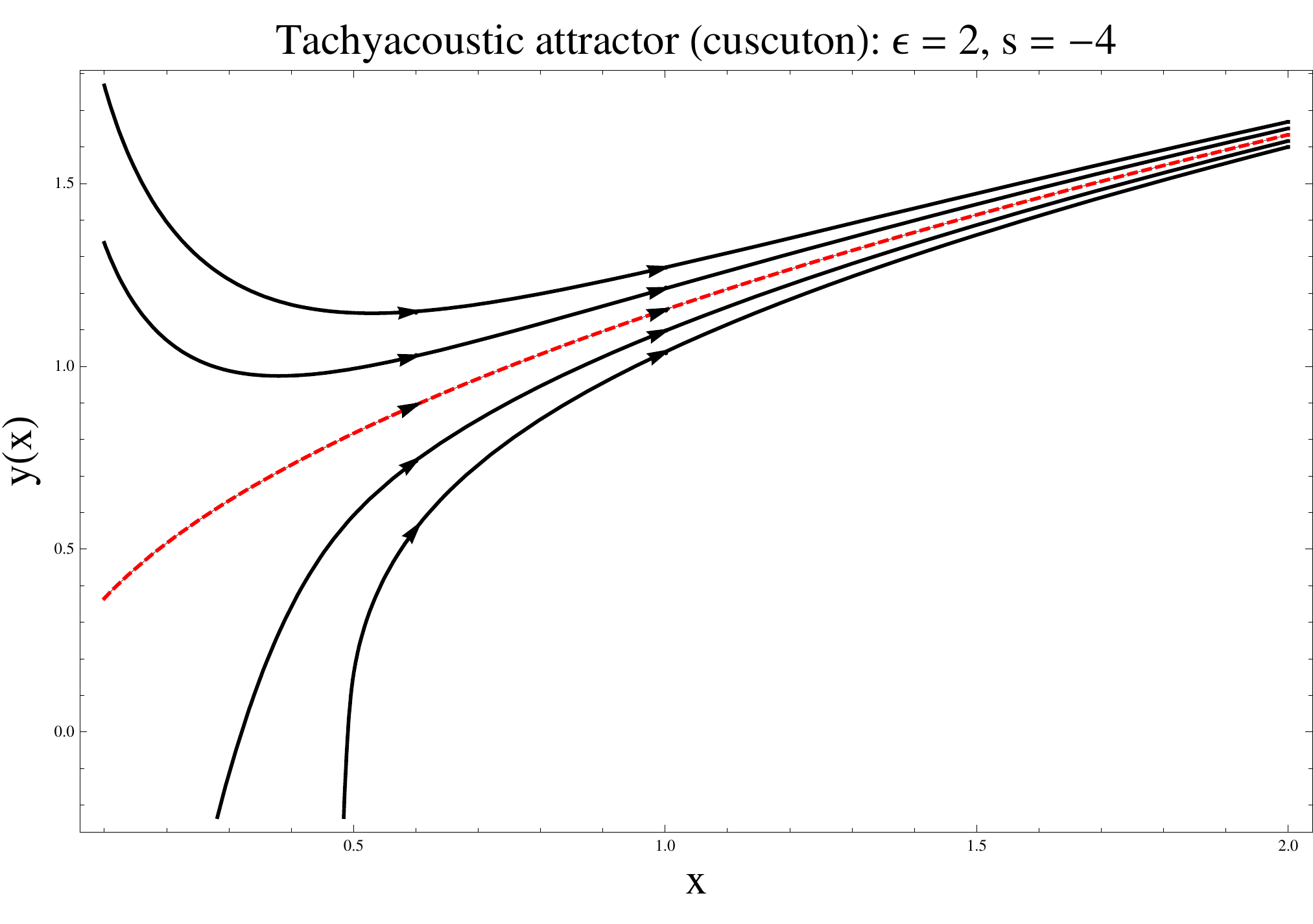}
\caption{Evolution of the dynamical variables $x$ and $y$ in a
radiation-dominated tachyacoustic model with a cuscuton Lagrangian. The Hamilton-Jacobi
trajectory is the red (dashed) line. \label{fig:cusc_rad}}
\end{figure}

\begin{figure}
\includegraphics[width=3.0in]{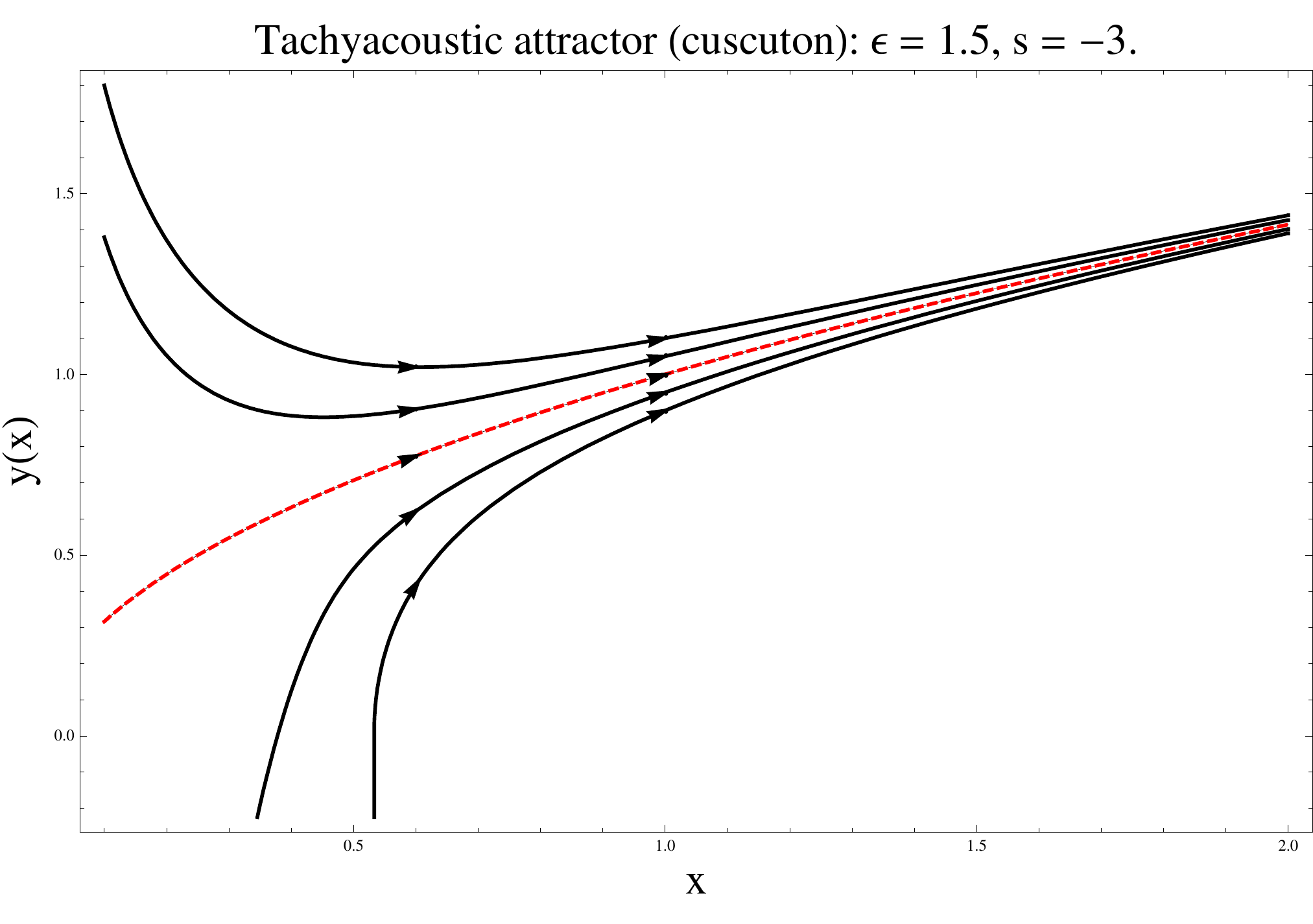}
\caption{Evolution of the dynamical variables $x$ and $y$ in a
matter-dominated tachyacoustic model with a cuscuton Lagrangian. The Hamilton-Jacobi
trajectory is the red (dashed) line. \label{fig:cusc_mat}}
\end{figure}

To study the attractor properties of the solution (\ref{eq:cuscutonsolution}), we construct a phase-space representation of the full equation of motion (\ref{eq:cusceom}). We introduce dimensionless phase space variables
\begin{eqnarray}
&&x \equiv \frac{\phi}{\phi_0},\cr
&&y \equiv \sqrt{\frac{2 \epsilon}{3 s^2}}\frac{\dot \phi}{H_0 \phi_0}.
\end{eqnarray}
The equation of motion can then be written as a phase-space evolution equation using \cite{mukhanov2005}
\begin{equation}
{\ddot\phi} = {\dot \phi}\frac{d {\dot\phi}}{d \phi} = \frac{3 s^2 H_0^2 \phi_0}{2 \epsilon} y\left(x\right) y'\left(x\right),
\end{equation}
so that
\begin{equation}
\label{cuscphase1}
y\left(x\right) y'\left(x\right) + 3 c_S^2 \sqrt{\frac{2 \epsilon}{3 s^2}} \frac{H}{H_0} y\left(x\right) + \frac{1}{3 M_P^2 H_0^2} V'\left(x\right) = 0.
\end{equation}
We define dimensionless versions of the potential $V$ and Hubble parameter $H$ as
\begin{equation}
v\left(x\right) \equiv \frac{V\left(\phi/\phi_0\right)}{3 M_P^2 H_0^2} = x^{2 \epsilon / s} \left(1 - \frac{\epsilon}{3} x^2\right),
\end{equation}
and
\begin{equation}
h\left(x\right) \equiv \sqrt{\frac{2 \epsilon}{3 s^2}} \frac{H}{H_0} = \sqrt{\frac{2 \epsilon}{3 s^2}\left(\frac{1}{2} y\left(x\right) + v\left(x\right)\right)},
\end{equation}
so that the phase space evolution equation (\ref{cuscphase1}) reduces to the simple form
\begin{equation}
\label{cusceom}
y\left(x\right) y'\left(x\right) + 3 c_S^2 h\left(x\right) y\left(x\right) + v'\left(x\right) = 0.
\end{equation}
Similarly, we define a dimensionless warp factor
\begin{equation}
g\left(x\right) \equiv \sqrt{\frac{2 \epsilon}{3}}\frac{s \phi_0}{\sqrt{2} M_P^2 H_0 \epsilon} f\left(\phi\right) = x^{\epsilon / s - 1}\left(1 - x^2\right).
\end{equation}
The sound speed is then
\begin{equation}
c_S^2 = 1 + \frac{g\left(x\right)}{y\left(x\right)}.
\end{equation}
In terms of the dimensionless variables, the analytic solution (\ref{eq:cuscutonsolution}) becomes
\begin{eqnarray}
\label{dimcuscsol}
&&y\left(x\right) = -s \sqrt{\frac{2 \epsilon}{3 s^2}} x^{\epsilon / s + 1},\cr
&&h\left(x\right) = \sqrt{\frac{2 \epsilon}{3 s^2}} x^{\epsilon / s},\cr
&&c_S\left(x\right) = \frac{1}{x}.
\end{eqnarray}
We evaluate the equation of motion (\ref{cusceom}) numerically to demonstrate that the solution (\ref{dimcuscsol}) is a phase-space attractor. Figure \ref{fig:cusc_rad} shows attractor behavior in the radiation-dominated case ($\epsilon = 2$), and Figure \ref{fig:cusc_mat} shows attractor behavior in the matter-dominated case ($\epsilon = 3/ 2$).

\subsection{The DBI Case}
\label{sec:attdbi}

Similar to the cuscuton case considered in the previous section,
we can construct a Dirac-Born-Infeld Lagrangian by choosing
${\tilde s} = s$, so that $n = 3$ in Eq. (\ref{difflag}), which
corresponds to the Lagrangian \cite{Bessada:2009ns}
\begin{equation}
{\mathcal L}\left(X,\phi\right) = -f^{-1}\left(\phi\right) \sqrt{1 - f\left(\phi\right) X} + f^{-1}\left(\phi\right) - V\left(\phi\right).
\end{equation}
For $\epsilon$ and $s$ constant we obtain an exactly solvable system, with
\begin{equation}
V\left(\phi\right) = 3 M_P^2 H_0^2 \left(\frac{\phi}{\phi_0}\right)^{-4 \epsilon / s} \left[1 - \left(\frac{2 \epsilon}{3}\right) \frac{1}{1 + \left(\phi/\phi_0\right)^2}\right],
\end{equation}
and
\begin{equation}
f\left(\phi\right) = \left(\frac{1}{2 M_p^2 H_0^2 \epsilon}\right) \left(\frac{\phi}{\phi_0}\right)^{4 \epsilon / s - 2} \left[1 - \left(\frac{\phi}{\phi_0}\right)^4\right].
\end{equation}
The sound speed (\ref{defspeedofsound}) for the DBI Lagrangian is
\begin{equation}
c_S = \frac{1}{{\mathcal L}_X} = \sqrt{1 - f\left(\phi\right) \dot\phi^2},
\end{equation}
and the Hubble parameter (\ref{eqFriedmann}) is
\begin{eqnarray}
H^2 &=& \frac{1}{3 M_p^2} \left(2 X {\mathcal L}_X - {\mathcal L}\right)\cr
&=& \frac{1}{3 M_p^2} \left[\frac{1 - c_S}{c_S f\left(\phi\right)} + V\left(\phi\right)\right].
\end{eqnarray}
The equation of motion for for the field $\phi$ is then
\begin{equation}
{\ddot\phi} + 3 H c_S^2 {\dot\phi} + \frac{3 f'\left(\phi\right)}{2 f\left(\phi\right)} {\dot\phi}^2 - \frac{f'\left(\phi\right)}{f\left(\phi\right)^2} + c_S^3 \left[\frac{f'\left(\phi\right)}{f\left(\phi\right)^2} + V'\left(\phi\right)\right] = 0,
\end{equation}
with solution
\begin{eqnarray}
&&\dot\phi = - \frac{s}{2} H_0 \phi_0 \left(\frac{\phi}{\phi_0}\right)^{-2 \epsilon / s + 1}, \cr
&&H\left(\phi\right) = H_0 \left(\frac{\phi}{\phi_0}\right)^{-2 \epsilon / s},\cr
&&c_S\left(\phi\right) = \left(\frac{\phi}{\phi_0}\right)^2.
\label{eq:DBIsol}
\end{eqnarray}
This solution again corresponds to power-law evolution $a \propto e^{-N} \propto t^{1/\epsilon}$. As in the cuscuton case, the spectral index of scalar perturbations is
\begin{equation}
n = 1 - \frac{2 \epsilon + s}{1 - \epsilon - s},
\end{equation}
so that the scale-invariant limit corresponds to $s = - 2 \epsilon$.

\begin{figure}
\includegraphics[width=3.0in]{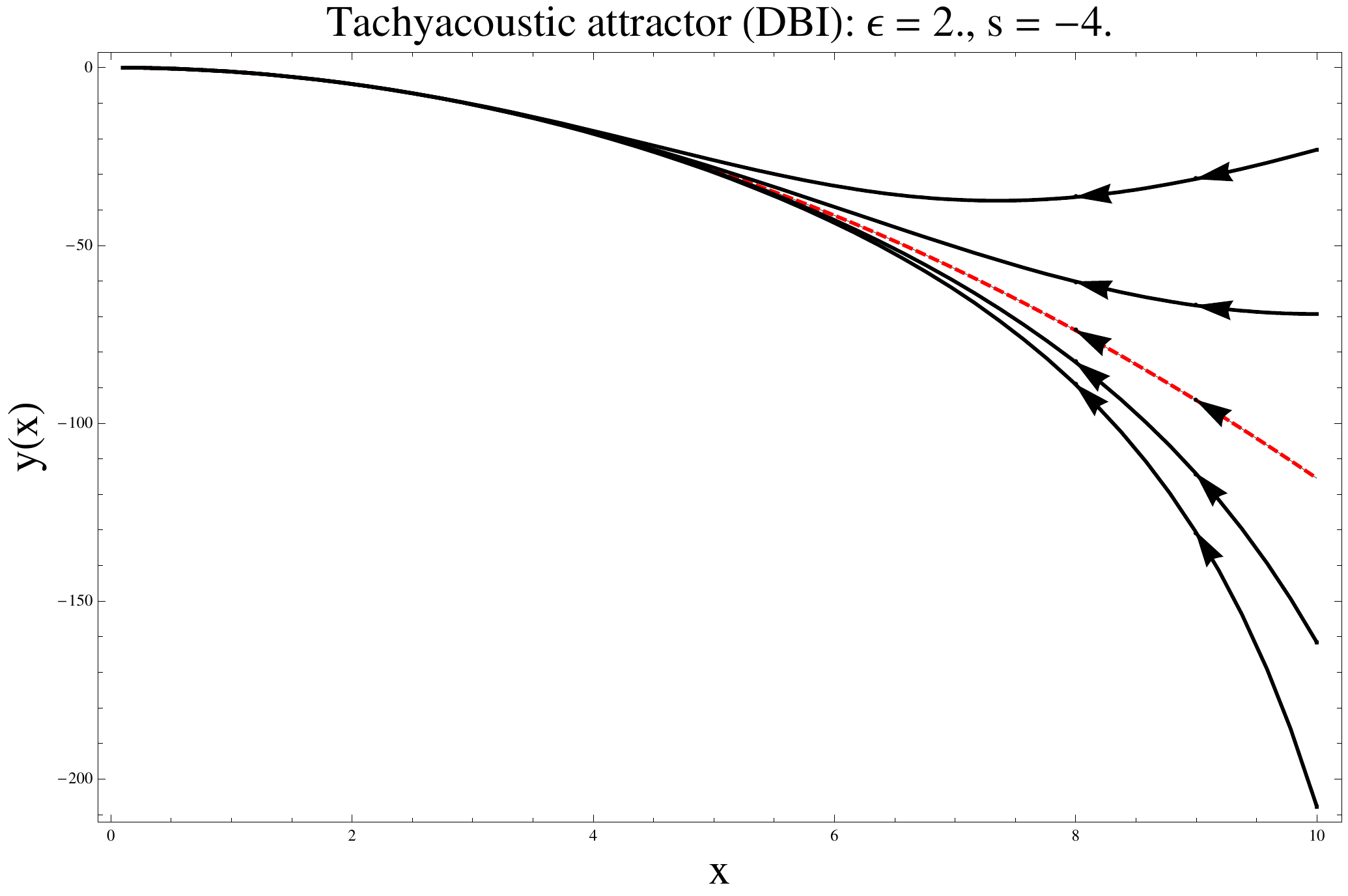}
\caption{Evolution of the dynamical variables $x$ and $y$ in a
radiation-dominated tachyacoustic model with a DBI Lagrangian. The Hamilton-Jacobi
trajectory is the red (dashed) line. \label{fig:dbi_rad}}
\end{figure}

\begin{figure}
\includegraphics[width=3.0in]{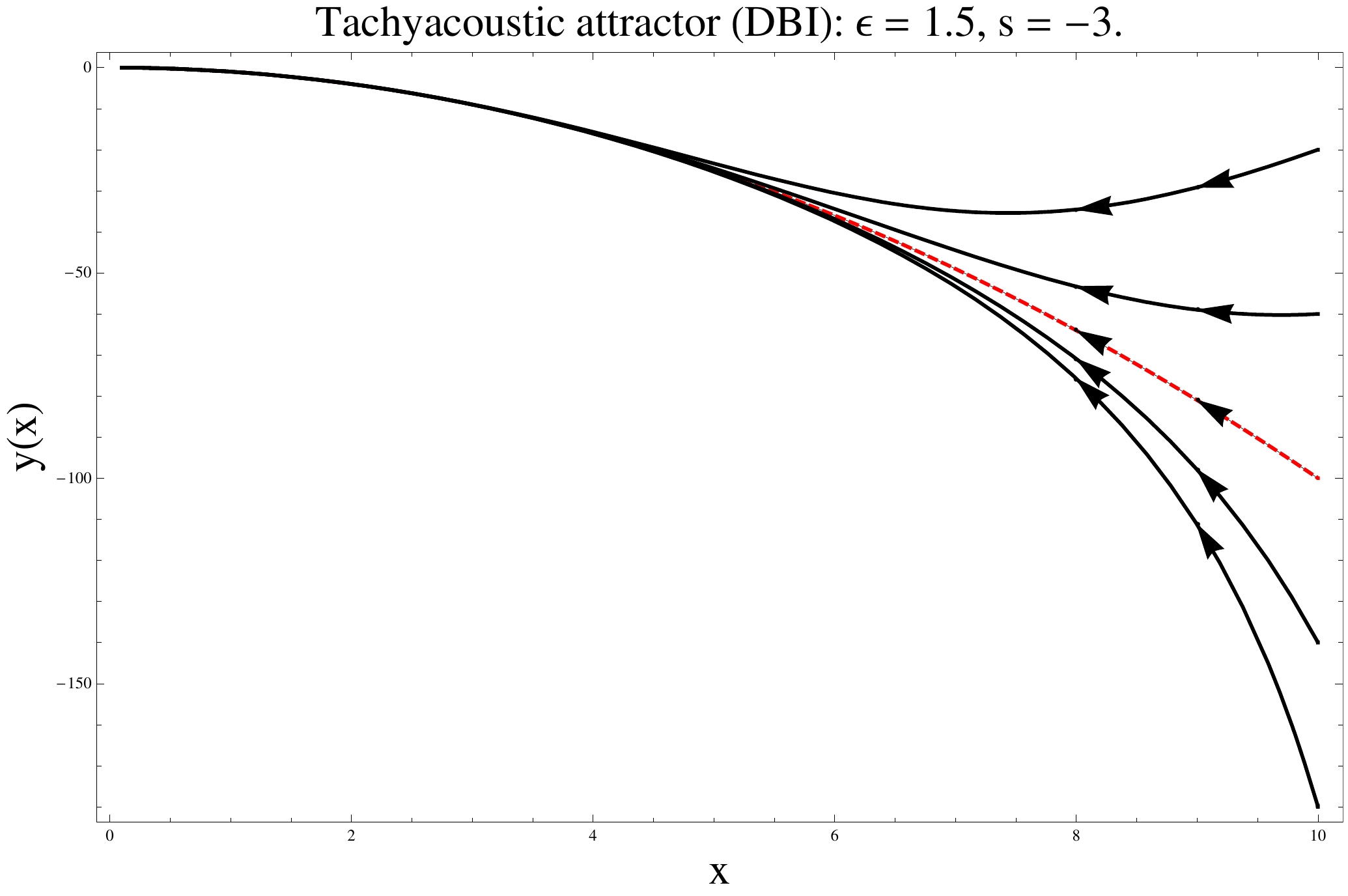}
\caption{Evolution of the dynamical variables $x$ and $y$ in a
matter-dominated tachyacoustic model with a DBI Lagrangian. The Hamilton-Jacobi
trajectory is the red (dashed) line. \label{fig:dbi_mat}}
\end{figure}

We study the attractor properties of this solution by defining dimensionless variables
\begin{eqnarray}
&&x \equiv \frac{\phi}{\phi_0},\cr
&&y \equiv \frac{\dot\phi}{\sqrt{3} H_0 M_p},
\end{eqnarray}
and
\begin{eqnarray}
&&v(x) \equiv \frac{V\left(\phi\right)}{3 M_p^2 H_0^2} = x^{-4 \epsilon / s}\left[1 - \left(\frac{2 \epsilon}{3}\right) \frac{1}{1 + x^2}\right],\cr
&&g(x) \equiv 3 M_P^2 H_0^2 f\left(\phi\right) = \left(\frac{3}{2 \epsilon}\right) x^{4 \epsilon / s - 2} \left(1 - x^4\right).
\end{eqnarray}
The dimensionless Hubble parameter is
\begin{equation}
h\left(x\right) \equiv \frac{H \phi_0}{\sqrt{3} M_P H_0} = \sqrt{\left(\frac{8 \epsilon}{3 s^2}\right) \left[\frac{1 - c_S\left(x\right)}{c_S\left(x\right) g\left(x\right)} + v\left(x\right)\right]},
\end{equation}
where the sound speed is
\begin{equation}
c_S^2 = 1 - g\left(x\right) y^2\left(x\right).
\end{equation}
In terms of these dimensionless variables, DBI equation of motion is:
\begin{equation}
y\left(x\right) y'\left(x\right) + 3 h c_S^2 y\left(x\right) + \frac{3}{2} \frac{g'}{g} y^2\left(x\right) + (c_S^3 - 1) \frac{g'}{g^2} + c_S^3 v' = 0.
\end{equation}
The analytic solution (\ref{eq:DBIsol}) is
\begin{eqnarray}
&&y\left(x\right) = \frac{s}{2} \sqrt{\frac{8 \epsilon}{3 s^2}} x^{1 - 2 \epsilon / s},\cr
&&h\left(x\right) = \sqrt{\frac{8 \epsilon}{3 s^2}} x^{-2 \epsilon / s},\cr
&&c_S = x^2.
\end{eqnarray}
We evaluate the attractor behavior of this solution in the scale-invariant limit $s = - 2 \epsilon$. Figure (\ref{fig:dbi_rad}) shows the radiation-dominated case ($\epsilon = 2$), and Figure (\ref{fig:dbi_mat}) shows the matter-dominated case ($\epsilon = 3/2$). In both cases, the analytic solution (\ref{eq:DBIsol}) is a dynamical attractor.

\section{Conclusions}
\label{sec:conc}

In this paper, we consider the dynamical stability of ``tachyacoustic'' cosmological models \cite{Bessada:2009ns}, which generate superhorizon cosmological perturbations via a decreasing, superluminal sound speed instead of accelerating expansion (as in the case of inflation). It is known that such cosmologies can produce nearly scale-invariant scalar perturbations, consistent with current data. However, it has not been previously shown that such solutions also correspond to dynamical attractors, which is a necessary condition for such models to be cosmologically viable. Such models are realized in scalar field theory via a non-canonical Lagrangian \cite{noncanonical}. In our analysis, we have considered two particular choices of Lagrangian which give rise to power-law evolution for the scale factor and sound speed, resulting in a scale-invariant primordial power spectrum. The first case is a so-called ``cuscuton'' Lagrangian, which is linear in the field kinetic term instead of quadratic as in the case of a canonical Lagrangian. We numerically integrate the full phase space for the field evolution, and show that the power-law tachyacoustic solution is in fact a dynamical attractor. Such models are of particular interest because they predict a detectable contribution to cosmological non-Gaussianity \cite{nongaussian}. Second, we consider a Dirac-Born-Infeld (DBI) Lagrangian giving identical power-law behavior and show that the power-law solution to this Lagrangian is also a dynamical attractor in the full phase space. We present results for the scale-invariant limit, but considering a slightly ``red'' spectrum as favored by data does not alter the attractor properties of the solution. We conclude that, like inflation, tachyacoustic cosmology can generate a scale-invariant power spectrum via dynamically stable cosmological evolution. We note that in this analysis we consider only the {\em classical} stability of tachyacoustic models. This may not apply at the quantum level. Negative-tension branes, for example, are known to have instabilities at the quantum level \cite{Nunes:2005up,Marolf:2001ne}. However, we are not aware of any general argument indicating that theories with a superluminal sound speed are necessarily unstable to quantum mechanical flucutations.

\begin{acknowledgments}
This  research is supported  in part by the National Science
Foundation under grant NSF-PHY-1066278. DB thanks the Brazilian
agency FAPESP, grant 2009/15612-6, for financial support at the
earlier stage of this work. WHK thanks the Kavli Institute for
Cosmological Physics at the University of Chicago, where part of
this work was completed, for generous hospitality.

\end{acknowledgments}


\begin{thebibliography}{99}


\bibitem{inflation}
   A.~H.~Guth,
   %``The Inflationary Universe: A Possible Solution To The Horizon And Flatness Problems,''
   Phys.\ Rev.\ D {\bf 23}, 347 (1981).
%%CITATION = PHRVA,D23,347;%%
   A.~D.~Linde,
   %``A New Inflationary Universe Scenario: A Possible Solution Of The Horizon, Flatness,
   %Homogeneity, Isotropy And Primordial Monopole Problems,''
   Phys.\ Lett.\ B {\bf 108}, 389 (1982).
   %%CITATION = PHLTA,B108,389;%%
   A.~Albrecht and P.~J.~Steinhardt,
   %``Cosmology For Grand Unified Theories With Radiatively Induced Symmetry Breaking,''
   Phys.\ Rev.\ Lett.\  {\bf 48}, 1220 (1982).
   %%CITATION = PRLTA,48,1220;%%

\bibitem{prebigbang}
G.~Veneziano,
  %``Scale Factor Duality For Classical And Quantum Strings,''
  Phys.\ Lett.\  B {\bf 265}, 287 (1991);
  M.~Gasperini and G.~Veneziano,
  %``Pre - big bang in string cosmology,''
  Astropart.\ Phys.\  {\bf 1}, 317 (1993)
  [arXiv:hep-th/9211021];
  M.~Gasperini and G.~Veneziano,
  %``The pre-big bang scenario in string cosmology,''
  Phys.\ Rept.\  {\bf 373}, 1 (2003)
  [arXiv:hep-th/0207130].
  V.~Bozza, M.~Gasperini, M.~Giovannini and G.~Veneziano,
  %``Assisting pre-big bang phenomenology through short-lived axions,''
  Phys.\ Lett.\  B {\bf 543}, 14 (2002)
  [arXiv:hep-ph/0206131];
  %%CITATION = PHLTA,B543,14;%%
V.~Bozza and G.~Veneziano,
  %``Regular two-component bouncing cosmologies and perturbations therein,''
  JCAP {\bf 0509}, 007 (2005)
  [arXiv:gr-qc/0506040].
  %%CITATION = JCAPA,0509,007;%%

\bibitem{ekpyrotic}
J.~Khoury, B.~A.~Ovrut, P.~J.~Steinhardt and N.~Turok,
  %``The ekpyrotic universe: Colliding branes and the origin of the hot big
  %bang,''
  Phys.\ Rev.\  D {\bf 64}, 123522 (2001)
  [arXiv:hep-th/0103239];
  %%CITATION = PHRVA,D64,123522;%%
P. Steinhardt and N. Turok, Science 296: 1436-1439, 2002.
J.~L.~Lehners, P.~McFadden, N.~Turok and P.~J.~Steinhardt,
  %``Generating ekpyrotic curvature perturbations before the big bang,''
  Phys.\ Rev.\  D {\bf 76}, 103501 (2007)
  [arXiv:hep-th/0702153];
  E.~I.~Buchbinder, J.~Khoury and B.~A.~Ovrut,
  %``New Ekpyrotic Cosmology,''
Phys.\ Rev.\  D {\bf 76}, 123503(2007) ÊÊÊ [arXiv:hep-th/0702154];
P.~Creminelli and L.~Senatore,
  %``A smooth bouncing cosmology with scale invariant spectrum,''
  JCAP {\bf 0711}, 010 (2007)
  [arXiv:hep-th/0702165];
  %%CITATION = JCAPA,0711,010;%%
  K.~Koyama and D.~Wands,
  %``Ekpyrotic collapse with multiple fields,''
  JCAP {\bf 0704}, 008 (2007)
  [arXiv:hep-th/0703040];
  %%CITATION = JCAPA,0704,008;%%
K.~Koyama, S.~Mizuno and D.~Wands,
  %``Curvature perturbations from ekpyrotic collapse with multiple fields,''
  Class.\ Quant.\ Grav.\  {\bf 24}, 3919 (2007)
  [arXiv:0704.1152 [hep-th]].
  %%CITATION = CQGRD,24,3919;%%

\bibitem{quantumcosmology}
J.~Acacio de Barros, N.~Pinto-Neto, and M.~A.~Sagioro-Leal, Phys.
Lett. A {\bf 241}, 229 (1998). R.~Colistete Jr., J.~C.~Fabris, and
N.~Pinto-Neto, \prd {\bf 62}, 083507 (2000). F.G. Alvarenga, J.C.
Fabris, N.A. Lemos and G.A. Monerat, Gen.Rel.Grav. {\bf 34}, 651
(2002).

\bibitem{noncanonical}
%\bibitem{ArmendarizPicon:1999rj}
  C.~Armendariz-Picon, T.~Damour and V.~F.~Mukhanov,
  %``k-Inflation,''
  Phys.\ Lett.\  B {\bf 458}, 209 (1999)
  [arXiv:hep-th/9904075].
  %%CITATION = PHLTA,B458,209;%%
%\bibitem{ArmendarizPicon:2006if}
   C.~Armendariz-Picon,
   %``Near scale invariance with modified dispersion relations,''
   JCAP {\bf 0610}, 010 (2006)
   [arXiv:astro-ph/0606168].
   %%CITATION = JCAPA,0610,010;%%
%\bibitem{Piao:2006ja}
  Y.~S.~Piao,
  %``Seeding of Primordial Perturbations During a Decelerated Expansion,''
  Phys.\ Rev.\  D {\bf 75}, 063517 (2007)
  [arXiv:gr-qc/0609071].
  %%CITATION = PHRVA,D75,063517;%%
%\bibitem{Magueijo:2008pm}
  J.~Magueijo,
  %``Speedy sound and cosmic structure,''
  Phys.\ Rev.\ Lett.\  {\bf 100}, 231302 (2008)
  [arXiv:0803.0859 [astro-ph]].
  %%CITATION = PRLTA,100,231302;%%
%\bibitem{Magueijo:2008sx}
  J.~Magueijo,
  %``Bimetric varying speed of light theories and primordial fluctuations,''
  Phys.\ Rev.\  D {\bf 79}, 043525 (2009)
  [arXiv:0807.1689 [gr-qc]].
  %%CITATION = PHRVA,D79,043525;%%
%\bibitem{Piao:2008ip}
  Y.~S.~Piao,
  %``On Primordial Density Perturbation and Decaying Speed of Sound,''
  arXiv:0807.3226 [gr-qc].
  %%CITATION = ARXIV:0807.3226;%%
%\bibitem{Piao:2004uq}
  Y.~-S.~Piao,
  %``On the dualities of primordial perturbation spectrums,''
  Phys.\ Lett.\ B {\bf 606}, 245 (2005)
  [hep-th/0404002].
  %%CITATION = HEP-TH/0404002;%%
%\bibitem{Piao:2011bz}
  Y.~-S.~Piao,
  %``Conformally Dual to Inflation,''
  arXiv:1112.3737 [hep-th].
  %%CITATION = ARXIV:1112.3737;%%


\bibitem{Geshnizjani:2011dk}
  G.~Geshnizjani, W.~H.~Kinney and A.~M.~Dizgah,
  %``General conditions for scale-invariant perturbations in an expanding universe,''
  JCAP {\bf 1111}, 049 (2011)
  [arXiv:1107.1241 [astro-ph.CO]].
  %%CITATION = ARXIV:1107.1241;%%


\bibitem{Bessada:2009ns}
  D.~Bessada, W.~H.~Kinney, D.~Stojkovic and J.~Wang,
  %``Tachyacoustic Cosmology: An Alternative to Inflation,''
  arXiv:0908.3898 [astro-ph.CO].

\bibitem{Kinney:2002qn}
  W.~H.~Kinney,
  %``Inflation: Flow, fixed points and observables to arbitrary order in  slow
  %roll,''
  Phys.\ Rev.\  D {\bf 66}, 083508 (2002)
  [arXiv:astro-ph/0206032].
  %%CITATION = PHRVA,D66,083508;%%

\bibitem{Bean:2008ga}
  R.~Bean, D.~J.~H.~Chung and G.~Geshnizjani,
  %``Reconstructing a general inflationary action,''
  Phys.\ Rev.\  D {\bf 78}, 023517 (2008)
  [arXiv:0801.0742 [astro-ph]].

\bibitem{Geshnizjani:2011rm}
  G.~Geshnizjani, W.~H.~Kinney and A.~M.~Dizgah,
  %``Horizon-preserving dualities and perturbations in non-canonical scalar field cosmologies,''
  JCAP {\bf 1202}, 015 (2012)
  [arXiv:1110.4640 [astro-ph.CO]].
  %%CITATION = ARXIV:1110.4640;%%

\bibitem{Kinney:2007ag}
  W.~H.~Kinney and K.~Tzirakis,
  %``Quantum modes in DBI inflation: exact solutions and constraints from vacuum
  %selection,''
  Phys.\ Rev.\  D {\bf 77}, 103517 (2008)
  [arXiv:0712.2043 [astro-ph]].

\bibitem{Afshordi:2006ad}
  N.~Afshordi, D.~J.~H.~Chung and G.~Geshnizjani,
  %``Cuscuton: A Causal Field Theory with an Infinite Speed of Sound,''
  Phys.\ Rev.\ D {\bf 75}, 083513 (2007)
  [hep-th/0609150].
  %%CITATION = HEP-TH/0609150;%%

\bibitem{Silverstein:2003hf}
  E.~Silverstein and D.~Tong,
  %``Scalar Speed Limits and Cosmology: Acceleration from D-cceleration,''
  Phys.\ Rev.\  D {\bf 70}, 103505 (2004)
  [arXiv:hep-th/0310221].
  %%CITATION = PHRVA,D70,103505;%%

\bibitem{Alishahiha:2004eh}
  M.~Alishahiha, E.~Silverstein and D.~Tong,
  %``DBI in the sky,''
  Phys.\ Rev.\ D {\bf 70}, 123505 (2004)
  [hep-th/0404084].
  %%CITATION = HEP-TH/0404084;%%

\bibitem{nongaussian}
  D.~Bessada,
  %``Non-Gaussian signatures of Tachyacoustic Cosmology,''
  JCAP {\bf 1209}, 018 (2012)
  [arXiv:1206.0728 [gr-qc]].
  %%CITATION = ARXIV:1206.0728;%%

\bibitem{Magueijo:2010zc}
  J.~Magueijo, J.~Noller and F.~Piazza,
  %``Bimetric structure formation: non-Gaussian predictions,''
  Phys.\ Rev.\ D {\bf 82}, 043521 (2010)
  [arXiv:1006.3216 [astro-ph.CO]].
  %%CITATION = ARXIV:1006.3216;%%

  \bibitem{Noller:2011hd}
  J.~Noller and J.~Magueijo,
  %``Non-Gaussianity in single field models without slow-roll,''
  Phys.\ Rev.\ D {\bf 83}, 103511 (2011)
  [arXiv:1102.0275 [astro-ph.CO]].
  %%CITATION = ARXIV:1102.0275;%%

\bibitem{Seery:2005wm}
  D.~Seery and J.~E.~Lidsey,
  %``Primordial non-Gaussianities in single field inflation,''
  JCAP {\bf 0506}, 003 (2005)
  [astro-ph/0503692].
  %%CITATION = ASTRO-PH/0503692;%%

\bibitem{Chen:2006nt}
  X.~Chen, M.~-x.~Huang, S.~Kachru and G.~Shiu,
  %``Observational signatures and non-Gaussianities of general single field inflation,''
  JCAP {\bf 0701}, 002 (2007)
  [hep-th/0605045].
  %%CITATION = HEP-TH/0605045;%%

\bibitem{Komatsu:2008hk}
  E.~Komatsu {\it et al.}  [WMAP Collaboration],
  %``Five-Year Wilkinson Microwave Anisotropy Probe (WMAP\altaffilmark 1 )
  %Observations:Cosmological Interpretation,''
  Astrophys.\ J.\ Suppl.\  {\bf 180}, 330 (2009)
  [arXiv:0803.0547 [astro-ph]].
  %%CITATION = APJSA,180,330;%%

\bibitem{mukhanov2005}
V.~Mukhanov,
``Physical Foundations of Cosmology'',
Cambridge University Press, Cambridge, UK (2005).

\bibitem{Nunes:2005up}
  N.~J.~Nunes and M.~Peloso,
  %``On the stability of field-theoretical regularizations of negative tension branes,''
  Phys.\ Lett.\ B {\bf 623}, 147 (2005)
  [hep-th/0506039].
  %%CITATION = HEP-TH/0506039;%%

\bibitem{Marolf:2001ne}
  D.~Marolf and M.~Trodden,
  %``Black holes and instabilities of negative tension branes,''
  Phys.\ Rev.\ D {\bf 64}, 065019 (2001)
  [hep-th/0102135].
  %%CITATION = HEP-TH/0102135;%%






\end{thebibliography}
\end{document}